%% file: ALEGRO18_rep_V0.tex
\documentclass{cernrep} 
\usepackage{texnames}
\usepackage[T1]{fontenc}
\usepackage[bookmarks, colorlinks=true, linktoc=page, pdftex, linkcolor=red, citecolor=red, urlcolor=red]{hyperref}
\begin{document}
\noindent \textbf{\LARGE{ALEGRO input  for the 2020 update of the European Strategy}} 

\vspace{0.25cm}
\noindent \textbf{\LARGE{for Particle Physics: comprehensive overview}}
\vspace{0.5cm}
 
\noindent \textbf{Contacts: B. Cros$^1$, P. Muggli$^2$ \\ 
on behalf of ALEGRO collaboration}, 

\noindent member list at 
http://www.lpgp.u-psud.fr/icfaana/alegro/alegro-members

\vspace{0.25cm}

\noindent{$^1$ LPGP, CNRS, Universit\'{e} Paris Sud, Orsay France, email: brigitte.cros@u-psud.fr\\
$^2$Max Planck Institute for Physics, Munich, Germany, email: muggli@mpp.mpg.de }

\vspace{1cm}


Advanced and Novel Accelerators (ANAs) can provide acceleration gradients orders of magnitude greater than conventional accelerator technologies, and hence they have the potential to provide a new generation of more compact, high-energy machines. Four technologies are of particular interest, all of which rely on the generation of a wakefield which contains intense electric fields suitable for particle acceleration. In the laser wakefield accelerator (LWFA) and plasma wakefield accelerator (PWFA) the wakefields are driven in a plasma by intense laser or particle beams, respectively; in the structure wakefield accelerator (SWFA), the wake is excited by a particle bunch propagating through a structured tube; and in the dielectric laser accelerator (DLA), a laser pulse directly drives an accelerating mode in a dielectric structure.

In view of the great promise of ANAs, and the substantial effort worldwide to develop them, the Advanced LinEar collider study GROup, ALEGRO, was formed at the initiative of the ICFA ANA panel. ALEGRO aims to foster studies on accelerators based on 
ANAs for applications to high-energy physics, with the ambition of proposing a machine that would address the future goals of particle physics. This document summarizes the current view of the international community on this topic.  It proposes a list of priorities that the community would like to invest effort in over the next five to ten years.

We propose as a long-term goal the design of an e$^+$/e$^-$/gamma collider with up to 30 TeV in the center of mass -  the Advanced Linear International Collider  (ALIC). On the path to this collider, a number of stepping stones have to be established. These will lead to spin-offs at lower energy that will benefit ultrafast X-ray science, medicine, and industrial applications. \textbf{The major goal for our community over the next five to ten years is the construction of dedicated ANA facilities that can reliably deliver high-quality, multi-GeV electron beams from a small number of stages. }The successful demonstration of robust stages of this type would provide a platform for 
ANAs with large number of stages generating high-quality beams in the TeV range.

The document also discusses other challenges that must be met for the complete ALIC concept. These include the design of appropriate particle sources, the development of high-power lasers needed for LWFAs and DLAs, the achievement of required tolerances, and the need for additional tools such as the development of novel diagnostics for the ultra-fast bunches generated by ANAs, and fast simulation methods.




 

\cleardoublepage

\section{Objectives of the ALEGRO input}

Following the outstanding progress made in Advanced and Novel Accelerators (ANAs)~\cite{cros2017} over the last ten years,
and the EU-funded networking activity, such as EuroNNAc \cite{euronnac},
the Advanced LinEar collider study GROup was formed at the initiative of the ICFA ANA-panel to foster studies on accelerators based on advanced accelerator concepts for applications to high-energy physics. 
ALEGRO members are physicists performing R\&D on ANAs,   representing a large, international community of researchers from universities and/or university groups, as well as various accelerator laboratories (CERN, DESY, INFN-LNF, PSI in Europe, 
ANL, LBNL, BNL, SLAC in the USA).
The ambition of ALEGRO is to propose a machine that would allow parameters meeting the requirements needed by particle physics to be reached. This document summarizes the current view of the international community on this topic, and \textbf{proposes a list of priorities that the community would like to invest effort in over the next five to ten years.}

ANAs are characterized by an accelerating gradient of the order of 1~GV/m and above, in plasmas or in vacuum dielectric structures. This makes them an attractive option for linear acceleration of relativistic leptons to high energy and potentially for application to an e-/e+ collider in the multi-TeV range. 
Such a machine, named \textbf{Advanced Linear International Collider (ALIC)},  would consist of an injector for each type of particle, many acceleration stages, a beam-delivery system and  detectors around the interaction point.

There are at least four acceleration techniques that could reach this average gradient: several options can be envisaged for ALIC, based either on a single technique, allowing the target parameters to be directly reached, or on a design combining several techniques. 
It is thus important to make progress with all techniques so that the most promising can emerge, or to determine the best combination of techniques.

A roadmap including the four techniques was proposed in 2017~\cite{cros2017} and ALEGRO's goal is to refine this roadmap and encourage collaborations  oriented towards the necessary R\&D that should allow us to propose a realistic machine design. 
The state of the art in advanced accelerators is such that it is  timely to propose global and ambitious designs that have all the characteristics of a collider, 
in terms of components and functions.

Although ALIC, the e-/e+ multi-TeV collider objective is on the time scale of 2037, several stepping stones have to be established along the way, leading to spin-offs  at lower energy for societal applications and contributing to our understanding of the physics processes underlying these novel techniques.
For example, the parameters of a 5 GeV plasma-based collider module would also be suitable for Free Electron Laser applications. Reaching control and feedback levels for reliable operation of accelerators in the context of collider development will have a tremendous impact on the dissemination of compact accelerators, even at lower energy, for application in medicine, biology, security and industry.

\section{Introduction to Advanced and Novel Accelerators}

The four ANAs considered are 
defined as follows: 
the \textbf{laser wakefield accelerator (LWFA)}, in which an intense laser pulse drives a
plasma wake; the beam-driven \textbf{plasma wakefield accelerator (PWFA)}, in which a particle bunch (e$^-$, e$^+$, p$^+$)
drives a plasma wake; the\textbf{ structure wakefield accelerator (SWFA),} in which a particle bunch drives wakefields in a structure, for instance a hollow dielectric tube or a corrugated
metallic tube; the \textbf{dielectric laser accelerator (DLA)}, in which a laser pulse directly drives an accelerating mode in a dielectric structure.

\begin{figure}[h]
	\begin{center}	
	\includegraphics[width=9cm]{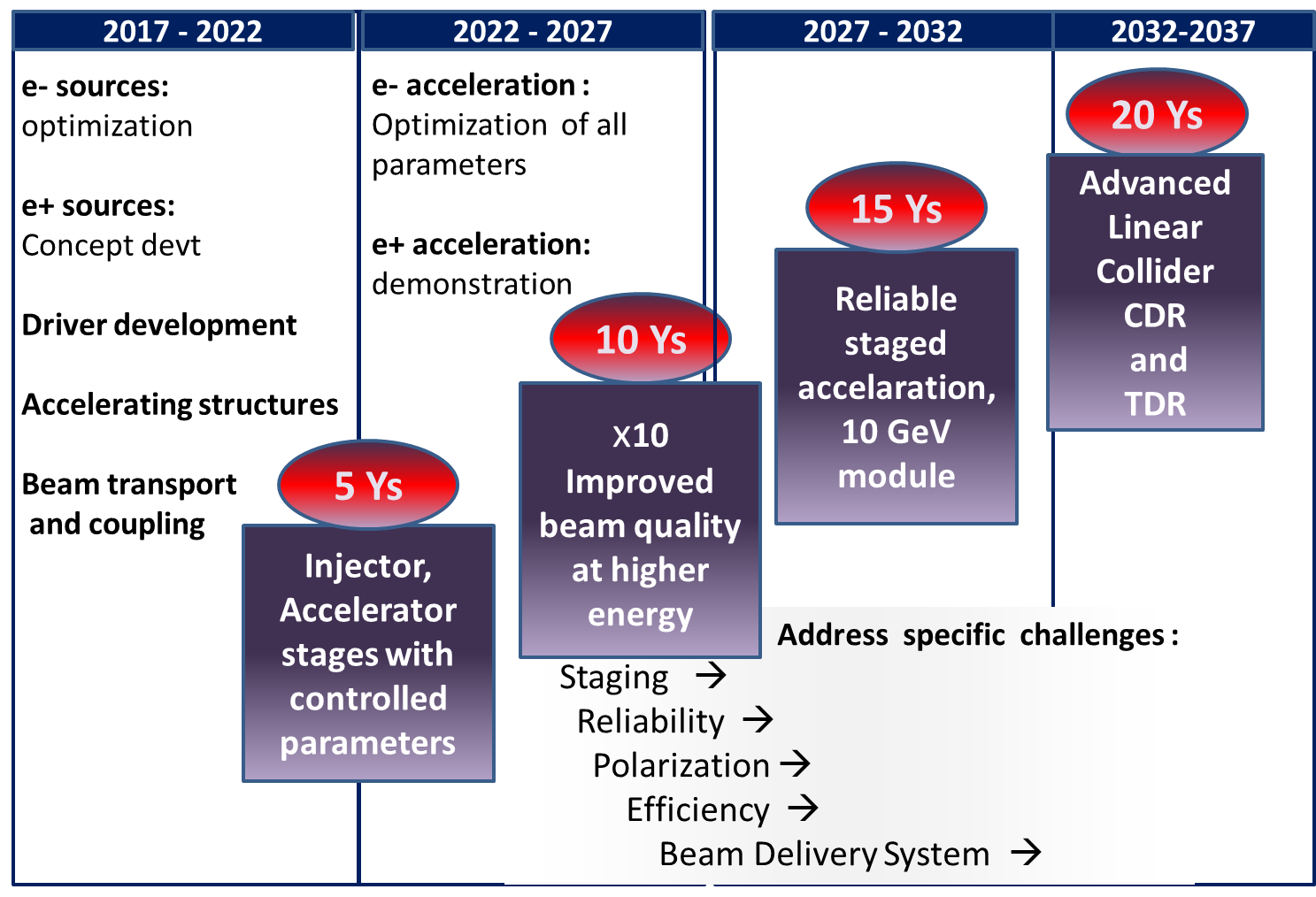}	
	\end{center}
    \caption{Global ANA scientific roadmap displaying phases and milestones towards the design of ALIC.}
    \label{roadmap}
\end{figure}

In ANAs using plasma as a medium (LWFA and PWFA), the wakefields are sustained
by a charge separation in the plasma and the accelerated bunch travels in the plasma. In
the ANAs using dielectric material as the medium,
the wakefields inside the vacuum core result from the boundary condition imposed by the material, and the accelerated bunch travels in vacuum.
 The ANAs scientific roadmap is shown in Fig.~\ref{roadmap}, highlighting the main phases expected towards ALIC. Details related to each technique can be found in Ref.~\cite{cros2017} and in the Addendum to this report.

\section{ Existing facilities for advanced acceleration R\&D}

\textbf{Existing facilities are mainly tools to explore ANA concepts, and proof-of-principle experiments. Dedicated facilities are needed to make significant progress in ANA R\&D towards the milestones in the next 5-10 years .}

LWFA R\&D uses PW-class peak power laser facilities. %
Most of these are multi-purpose, multi-user facilities with very few dedicated beam lines for accelerator R\&D.
Therefore, most of these lasers systems have state-of-the-art components and performance, with repetition rate lower than 1~Hz. 
These types of facilities have demonstrated many of the elements important for collider designs, though not necessarily at the same time: 5 GeV electron acceleration over 9 cm, high-quality injector, staging with independent laser beams.
The stability and reproducibility of laser and electron beams are limited by the low investment level into these research topics. One of the main applications driving the development of the field is the generation of compact light sources. 
The provision of high-power laser facilities with beamlines dedicated to ANA research is essential to drive the required  increased rate of progress in addressing key challenges in issues such as controlled injection, staging, or improved stability.


There are a limited number of facilities around the world suitable for PWFA experiments, fewer than ten, due to the small number of accelerator facilities that can deliver suitable particle bunches: GeV energy, nC charge in sub-ps length and focused to sub-mm transverse size. Similar to the LWFA case, facilities have demonstrated many of the elements important for collider designs, though not necessarily at the same time: 43~GeV energy gain in an 85~cm long plasma, beam loaded acceleration with percent level energy spread and high-efficiency. The majority of R\&D focuses on the acceleration of electrons, since positrons bunches are only available at one facility (FACET-II), but important aspects of positron acceleration have also been demonstrated: multi-GeV acceleration in meter scale plasmas and two-bunch acceleration in hollow-channel plasma structures. Acceleration with simultaneous beam loading, narrow relative energy spread and emittance preservation is currently being addressed. No facility can currently deliver independently shaped drive and witness bunches necessary to demonstrate maximum efficiency. Two main applications are under investigation: electron beam-driven PWFA with possible application to free electron lasers and e$^-$/e$^+$ collider; and proton driven PWFA, with possible application to beam-dump experiments, dark-matter searches and e$^-$/p$^+$ collisions.

Current SWFA R\&D focuses on investigating efficient structures capable of reliable operation at high gradients. In addition, SWFA requires drive- and main-bunch development. Controlling drive-bunch distributions provides a path to achieving high-gradient high-efficiency operation while controlling the main-bunch parameters optimises luminosity and efficiency.
R\&D focuses on achieving large-transformer-ratio acceleration with appropriately shaped bunches, which can achieve higher energy, with higher efficiency, than with unshaped bunches. %

The DLA scheme greatly benefits from progress in the telecommunication and semi-conductor industry both with high-repetition-rate fiber laser systems and with micro-fabrication. DLA is currently being studied on a compact scale compatible with university laboratory experiments. Relativistic-energy experiments are also being carried out at newly commissioned test facilities.
DLA has unique requirements for low-emittance electron sources, so facilities with suitable beam parameters are needed in order to test DLAs with linear-collider-type parameters.

 \section{Physics case}

\input{shortphysicsforALEGRO}

\section{Our long-term goal: ALIC, the Advanced Linear International Collider}

 To reach 30 TeV center-of-mass energies, a lepton collider based on traditional RF microwave technology would need to be over 100 km in length and would likely cost tens of billions of dollars to build.  Due to the inverse scaling of the interaction cross section with energy, the required luminosity  goal would be  of the order of $10^{36}$ cm$^{-2}$s$^{-1}$.  The design of ALIC will attempt to meet these requirements in a smaller cost/size footprint using advanced acceleration schemes.
 
 ANAs promise an important increase of accelerating gradient compared to existing collider technologies, but with different relative advantages and development needs. Among these the promise in gradient is highest for plasma techniques (LWFA, PWFA), while laser driven ANAs (LWFA, DLA) are expected to have the most compact footprint. LWFA development will rely on innovation in laser technology to improve wall-plug efficiency, while laser requirements for DLA are largely met by existing laser technology. The SWFA is conceptually similar to the relatively mature CLIC design but operates with higher gradient.

The current effort on acceleration technology should be complemented by substantial R\& D efforts on establishing concepts for colliders.
 \textbf{To date each advanced technology is outlining a design of collider based on its technology}, as reported in the Addendum~\cite{Addendum}.
\textbf{The main role of ALEGRO will be to identify the technology best suited for each collider component and at least one realistic global design.}
ALIC will consist of many successive modules or stages, with designs to reach 0.25, 1 and 30 TeV.  Many collider components - e.g. beam damping and cooling, beam-delivery systems, final focus, etc. -  are in the conceptual stage, or have not been considered in detail.  The goal of the R\&D will be to explore accelerator physics and collider components to enable a detailed design.

\section{ALIC Machine components}


\subsection{Electron/positron sources, cooling}

\textbf{The development of high-quality electron and positron sources is a high priority for ANAs.}

Compact electron sources based on laser-assisted field emission from nanotips, which can produce electron beams of unprecedented brightness, are under development for the DLA. High-gradient, high-energy superconducting radio-frequency (SRF) guns operated in continuous wave (CW) mode are promising candidates for delivering relevant beams for DLA-based linear colliders.
Comparable techniques for generating high-brightness positron beams in a compact footprint have not yet been identified. %

The SWFA and the PWFA use beams similar to those generated for CLIC or ILC. %
One can therefore take advantage of these developments and adapt them for the specific parameters required by the SWFA and the PWFA (shorter bunch, repetition rate, etc.). %

Plasma electrons can be trapped by wakefields, emerging in short bunches and possibly with very low emittance, at the nm-rad level. %
These would not require further cooling in a damping ring and could thus significantly contribute to the simplification of the collider design. %
Generation of beams with emittance at the level required for a collider (emittance, charge, etc.) must be demonstrated experimentally. %

The LWFA technique can in principle be free of any conventional accelerator components. %
Therefore, a number of promising plasma-electron injection schemes are studied. %
A short (few mm), LWFA stage could be specifically designed as an injector. %
Many encouraging results have been obtained experimentally. %
However, producing collider-like electron bunches remains a challenge. %

At the present time, positron sources envisaged would be similar to those already existing, i.e., using gamma rays and pair generation in a high-Z target, followed by capture and cooling in a damping ring. %
These were developed for ILC and CLIC and would have to be adjusted for the ANA beam format. %

\subsection{Accelerating structures}

The accelerating structure is the best developed component of the collider system. %
With all four ANAs, acceleration with GV/m gradient has been demonstrated experimentally. %
Multi-GeV energy gains have been obtained in the LWFA and the PWFA. %
\textbf{Challenges on the road to ALIC include beam-quality generation and preservation, energy transfer efficiency and operation with collider-like parameters.} %

The DLA accelerating structure consists of many acceleration modules synchronized to one laser oscillator. %
It is very similar to a metallic structure system, but the overall system is scaled down in wavelength from 10~cm to order of 1~$\mu$m. %
The typical structures under current active experimental study are periodic planar structures, where the power is delivered by laser pulses coupled sideways through the dielectric directly into the accelerating structure. DLA waveguide structures where the laser pulse co-propagates with the electron beam have been developed and may be more suitable for linear-collider applications.
The structure can accelerate electrons as well as positrons.  %

Dielectrics materials have higher damage threshold than metals. %
There are two SWFA configurations: collinear-wakefield acceleration (CWA) requiring only one beamline, and two-beam acceleration (TBA) requiring two beam lines (similar to CLIC). %
In the CWA configuration, the accelerating structure consists of a continuous hollow dielectric tube enclosed in a metallic cladding. %
Wakefields are generated by Cherenkov radiation of the drive bunch fields into the dielectric that are reflected towards the axis by the cladding. %
In the TBA configuration, the decelerating and accelerating structures consist of a metallic tube with dielectric irises placed periodically (similar to a pillbox structure). %
The CWA operate at higher frequencies (hundreds of GHz) and generally support higher gradient than the  TBA operating at lower frequencies (tens of GHz). %

In the LWFA and the PWFA, the accelerating structure consists of a periodic and axially symmetric perturbation in the plasma electron density. %
The structure can sustain very large (>100~GV/m) peak fields, both longitudinal (accelerating/decelerating) and transverse (focusing/defocusing). %
The fields are excited by an intense laser pulse or a relativistic, dense, charged-particle bunch (e$^-$, p$^+$). %
Typical electron plasma densities are in the 10$^{14}$ - 10$^{17}$cm$^{-3}$ range. %

The quality of the accelerating structure, and thus of the main beam, depends on the characteristics of the plasma and on that of the driver. %
Plasma sources that have suitable density uniformity and density reproducibility to minimize the bunch relative energy spread must be developed. %
Sources capable of operating at relatively high repetition rate (thousands of times per second) and for long periods at a time (months) must be developed. %
Plasma sources used in current demonstrations 
are capillary discharges, gas cells, electrical discharges and alkali-metal vapor sources. %
The LWFA requires laser-pulse guiding to extend the acceleration distance beyond the Rayleigh length of the laser beam. %
Sources longer than 30\,cm exist. %
For the PWFA, metal vapor sources exceeding one meter in length exist and strong plasma focusing provides guiding of both the drive and main beams. %

\subsection{Coupling/transport components between stages} 

Transport of the main beam between high-gradient acceleration stages is critical. %
\textbf{To maximize the average accelerating gradient and thus minimize the length of the accelerator, the distance between accelerating structures must be minimized. %
}

For DLA, focusing can be done by designing structures to provide an inherent focusing force from the laser field. Photonic structures for guiding and coupling the laser light to the DLA accelerator are also being actively developed.

Considering the similarity between the TBA configuration of SWFA and already designed colliders (CLIC, ILC), coupling between SWFA stages will be similar to that used in these collider designs. However, the CWA presents challenges to preserve main-beam quality during the staging. %

LWFAs and PWFAs are characterized by very strong focusing in the accelerating structure. %
Density tapering at the plasma entrance and exit of each accelerating structure is used to smooth the transition between strong plasma focusing and weak focusing of the magnetic optics used to capture the main beam from the previous structure and re-focus it into the next. %
This is an important research topic also aimed at minimizing the distance and the emittance growth between stages. %

Active plasma lenses also operate in the strong focusing regime, have axially symmetric focusing, unlike magnetic quadrupoles, and work equally well with e$^-$ and e$^+$.
Their use could then significantly reduce inter-stage length and emittance growth. %
However, their effectiveness for beams with the density envisaged for collider beams needs to be confirmed. %

In the case of the PWFA and colinear SWFA, inter-stage optics must also accommodate in- and out-coupling of the drive bunch. %
This is particularly challenging for the first few stages of the collider, where drive- and main-beam energies are similar. %
Further detailed studies, in particular on the preservation of the main-beam emittance must be performed. %
In- and out-coupling of drive laser pulses with the LWFA is greatly simplified by the use for example of plasma mirrors to bend the drive-pulse photons in and out of the path of the main beam over very short distances (<1\,m). %

\subsection{Drivers }

The requirements of the driver depend critically on the wakefield scheme. Multiple drivers are required for staging:  experiments using at least two drivers and accelerating structures should be performed to demonstrate the scalability of the techniques to reach high energies. %


The requirements of the drive lasers  needed for a DLA-based accelerator reflect the unusual pulse format of the electron beam, namely: pulse energies in the range of 1 to 10$\mu$J,  pulse duration below 1 ps, pulse repetition rates of 10 to 100 MHz, and high (>30\%) wall-plug efficiencies.
It is likely that fiber laser will be used to drive DLAs since they are efficient, robust, and low maintenance operation. %
\textbf{The  state of the art in laser systems is not far from what will  be required for large-scale DLA-based accelerators.}


More significant progress in laser technology will be required for application of LWFAs to ALIC. LWFAs providing energy gain close to 10 GeV have been demonstrated, using chirped pulse amplification (CPA) and titanium-doped sapphire lasers pumped by flashlamp-pumped lasers. However, these lasers have low wall-plug efficiency and, at the pulse energies required, are limited to repetition rates of a few hertz. Fortunately, substantial collaborative efforts between research laboratories and industry are underway to develop efficient multi-joule, diode-pumped short-pulse lasers operating at kHz repetition rates and beyond, with several options being pursued.  Longer wavelength operation of some of these options presents additional design space over which to evaluate and optimize LWFA system performance.

\textbf{Most lasers used today for LWFA are forefront technology, research instruments, and hence they often lack the control and feedback systems that will be required to generate stable particle beams.} The development of suitable feedback systems is an area of increasing activity; for example, it is a major component of the EuPRAXIA project. Future increases in laser repetition rates will have ancillary benefits from the perspective of active feedback-control of laser performance.


The CWA-SWFA and the PWFA have similar drive-beam requirements. 
This applies both to drive and witness beams, though plasma-based sources may provide ultra-low-emittance main beams without damping ring. %
Since both acceleration techniques operate at higher frequencies than CLIC, the system must provide shorter bunches. %
Therefore, an XFEL type beam is an excellent driver. %
The PWFA requires a higher-energy drive beam ($\sim$25 GeV). %

In the SWFA, drive- and main-bunch shaping can be effectively used to increase the transformer ratio and the energy-transfer efficiency. %
Bunch shaping has been demonstrated experimentally with various techniques. %
Acceleration with large transformer ratio ($\sim4$) has been demonstrated experimentally. %
Non-destructive techniques that can precisely shape bunches reaching the accelerating structure entrance must be developed and demonstrated. %
Numerical simulations show that bunch shaping can also greatly improve the energy-transfer efficiency in the PWFA. %

An essential element of the driver system is the distribution of the bunches to the successive acceleration stages. %
In the TBA scheme, this system was developed for CLIC and can in principle be re-used, with appropriate modifications for the SWFA specific parameters. %

The use of p$^+$ bunches to drive wakefields is also under study. %
Proton bunches carry large amounts of energy (tens to hundreds of kilojoules) to accelerate a main beam to the TeV energy range in a single plasma, effectively alleviating inter-stage issues. %
However, available p$^+$ bunch repetition rates limit the application of this scheme to an e$^-$/p$^+$ collider that can do interesting physics with lower luminosity than an e$^-$/e$^+$ collider. %

\subsection{Beam Delivery System: IP components and detectors }

The ALIC BDS and detector must be designed to take into account the particular format of the beams produced by DLA, LWFA and PWFA accelerators.  The various ANA methods favor different formats, from bunch trains similar to those of CLIC for LWFA and PWFA to small bunches with sub-ps  spacing for DLA. These different formats also lead to different levels of beam-related background from e$^-$/e$^+$ pairs. The detector design should accommodate these.  A sample detector model is presented in the Addendum~\cite{Addendum}. In particular, the ultra-short bunches, short when compared to the beamstrahlung formation length (beam collisions in the highly quantum beamstrahlung regime) could allow high luminosity to be reached with round beams~\cite{yaki}.

\section{Integrated system}

Various aspects of system integration specific to ANAs need to performed.
New schemes for coupling of beams  between stages, such as   plasma lens or plasma mirrors are promising concepts to maintain the average gradient of plasma accelerators. %
Their impact on the beam properties and the possibility to use them at high energy need to be evaluated.

\subsection{Tolerances}

ANAs naturally operate with very short and very small bunches. Temporal tolerances in terms of drive- and main-beam alignment must be correspondingly precise.  Spatial tolerances are driven by the small transverse size of the accelerating structures. One can anticipate that these temporal and spatial tolerance requirements throughout the accelerator are similar to those in conventional machines when expressed as a fraction of the size of the bunches and accelerating cavities.
The impact of these bunch parameters on the tolerances specified for a collider interaction region must also be evaluated.
In the case of LWFA and PWFA, tolerance studies require lengthy, systematic numerical simulations in 2 or 3D. Therefore, more effective numerical methods capable of simulating individual or series of stages must be developed.
These studies have not started since most current devices operate in a research mode rather than a beam-production mode. However, \textbf{tolerance studies are an important aspect of a collider design and such studies must be encouraged and performed.}

\subsection{Instrumentation}

ANAs produce bunches that are shorter than those of planned colliders (ILC, CLIC). %
Most standard beam diagnostics can be used, providing they operate with short bunches. %
New and suitable diagnostics will be developed as the need arises at demonstration facilities. %

\subsection{Simulations}

The required level of design effort for a collider will drive development of enhanced simulation capabilities. With sustained support of simulation efforts, it is envisioned that computer hardware and software developments will reduce the time frame for detailed simulations of one collider stage from weeks to hours or even minutes on future exascale-capable supercomputers. %
 
The workhorse algorithm for modeling ANA concepts is the Particle-In-Cell (PIC) methodology.  
High-resolution, full three-dimensional standard PIC and quasistatic PIC 
simulations are ultimately needed to capture potential hosing, misalignments, tilts and other non-ideal effects. Ensemble runs of simulations over large parameter space are required to estimate tolerances to those effects as well as study various designs. 
Hence, it is essential to pursue the development of better algorithms that improve the accuracy of existing plasma physics models (e.g. high-order Maxwell solvers, adaptive time-stepped particle pushers, adaptive mesh refinement, control of numerical instabilities) as well as to port the codes to the next generation of massively parallel supercomputers with multi-level parallelism. %
Interoperability of codes on the various computer architectures will be essential to reduce the cost and complexity of code development and maintenance.

In addition to the detailed PIC-based models, it is also important to develop fast tools that require far less computational resources and  that can be used to guide the parameter scans. The models used in these tools will be guided by theory and fits to the PIC-based simulations and experimental results. For the latter, the ANA community should start to take advantage of Artificial Intelligence (AI) - Machine Learning (ML) and develop models based on accumulated data from simulations and experiments with advanced AI-ML algorithms. 

For PWFA/LWFA, the priorities for the modeling of the plasma physics for a single stage will be to (a) demonstrate that reduced models can be used to make accurate predictions for a single stage producing beams in the range between 1 GeV and 1 TeV (in the linear, mildly nonlinear and nonlinear regimes), (b) demonstrate that the PIC algorithm can systematically reproduce current experimental results (coordinated efforts with experiments) and (c) study tolerances and the role of misalignments. It is also essential to integrate physics models beyond single-stage plasma physics in the current numerical tools, incorporating: plasma hydrodynamics for long-term plasma/gas evolution and accurate plasma/gas profiles, ionisation \& recombination, coupling with conventional beam lines, production of secondary particles, QED physics at the interaction point, and spin polarization. For DLA/SWFA, the priorities for the development of physics models for 3D  structures will be to (a) determine the material response to 
 electron bunches that are tens of attosecond long, (b) develop computational models based on material science experiments, (c) include material erosion into available models and (d) establish automated and non-invasive mechanisms to determine which photonics structures are useful for acceleration after being produced. %
For SWFA, the modeling needs include the development of simulation tools capable of precisely handling the electromagnetic (e.g. dispersion) and physical (e.g. response to strong field) properties of the considered materials. 

In addition, the tools and expertise developed for conventional accelerators will be exploited. %

\section{What is needed? What do we support?}

The R\&D during the upcoming years will focus on producing quality accelerated bunches (emittance and energy spread) with higher and higher energies (multi GeVs). %
\textbf{The results of this R\&D depend on the availability of suitable laser and particle beams.} %
Such R\&D often requires large facilities that can deliver laser pulses of hundreds of terawatt to petawatt power, or multi-GeV electron bunches to drive wakefields in plasmas (LWFA and PWFA) or dielectric structures (SWFA). %
The results also strongly depend on the level of control of the experimental parameters. %
\textbf{The exquisite control needed to reach collider parameters comes at the cost of a large investment in feedback and control systems.} %
All these elements explain why much of the progress towards collider development is expected to occur at facilities hosted by national laboratories, in synergy with contributions from smaller, university groups playing a key role through the generation of new ideas and concepts.

\textbf{ A number of key topics related to what could be the first stage  of ALIC, consisting of an injector plus accelerator module, and  producing beams in the 5-25 GeV range, are planned to be addressed by the international community at running or upcoming facilities, } as listed below:
\begin{itemize}
\item[$\bullet$] \textbf{External injection} of a quality electron bunch in an accelerator section will be studied at all the facilities of Table~\ref{tab:facilities}. %
The main-beam source may be produced by a conventional accelerator (RF-photo-injector gun) or a plasma-based injector. %
Emittance at the mm-mrad level should be produced and maintained, while the final energy spread should be at the percent or lower level. %
Low energy spread is reached by beam loading and bunch shaping will be used to increase energy-transfer efficiency and energy gain. %

\item[$\bullet$]
For collider applications, it is essential that ANAs reach levels of \textbf{bunch quality, efficiency, stability and reproducibility} equivalent to those produced by conventional accelerators. %
This is the main goal of the design study EuPRAXIA.  %
Reaching beam quality sufficient to reach self-amplified spontaneous emission (SASE) in a wiggler with a beam in the 1-5 GeV energy range is the main focus of
FlashFWD at DESY and of the design study EuPRAXIA. 
Dedicated test areas for DLA experiments now under construction at DESY and PSI will allow injection and transport of high-quality bunched beams and higher injected beam powers to be studied.%

\item[$\bullet$]
\textbf{Plasma sources }with sufficient density uniformity, control and reproducibility of the density, as well as tapering of the entrance and exit density ramps to assist in beam manipulation between stages must be developed. %
 Very high energy gain can in principle be reached in a single, long plasma stage by using a proton bunch carrying a large amount of energy as a driver. %
 This is the main thrust of the AWAKE experiment at CERN. %

\item[$\bullet$]
\textbf{Operation at high repetition rate} is an important step towards  ALIC. %
Operation at a kilohertz repetition rate requires development of lasers that must therefore produce kilo to mega-watts of average power. %
This challenge will be addressed by the kBella project at LBNL. %
Operation at high repetition rate also brings challenges for the plasma source, in particular for active ones energized by a pulsed electrical circuit and for those with walls close to the drive beam. %
This challenge will be looked at by several groups.

\item[$\bullet$]
Challenges in producing high-quality electron (e$^-$) and \textbf{positron (e$^+$) bunches} in the 10 to 20 GeV energy range will be addressed at FACET II at SLAC. %
We emphasize here that, in the near future, e$^+$ bunches suitable for LWFA or PWFA experiments will only be available at FACET II. %
\textbf{It is therefore imperative that e$^+$ bunches become available at other facilities, in particular in Europe. }%
This could be possible at CERN, from LEP components, or at facilities such as EuPRAXIA using novel methods to produce e$^+$ bunches. %

\item[$\bullet$]
The availability of \textbf{independently shaped drive- and main-beam} PWFA facilities would provide the ability to control the bunch parameters with sufficient accuracy and tune experiments to reach collider-level beam parameters, in particular in terms of beam loading and energy spread. %
Facilities  in the planning phase, such as at MAX IV in Lund Sweden or the eSPS at CERN may have this ability. %

\item[$\bullet$]
Dielectric structures of the SWFA are ready for extensive \textbf{breakdown} tests to determine the operational accelerating gradient under collider conditions, i.e. with breakdown rates in the 10$^{-7}$ range. %
These structures, operating at higher gradients than those of CLIC, could replace the CLIC structures or could be considered for the CLIC energy upgrade. %
Breakdown tests and investigations of the integration of the structure in a CLIC scheme could be performed in a collaboration between Argonne AWA and CERN. %

\end{itemize}

\begin{table}[h]
\caption{Facilities for accelerator R\&D in the multi-GeV range relevant for ALIC and with emphasis on specific challenges}
\label{tab:facilities}
\centering
\begin{tabular}{lllll}\hline\hline
Facility & Readiness  & ANA technique & Specific Goal 
\\
&&&  
\\ \hline
kBELLA  & Design study & LWFA  & e-, 10 GeV, KHz rep rate\\
EuPRAXIA & Design study & LWFA or PWFA & e-, 5 GeV, reliability \\
AWAKE & Operating & PWFA & e$^-$/p$^+$ collider\\
FACET II  & Start 2020 & PWFA & e$^-$, 10 GeV boost, beam quality, e$^+$ acceleration \\
Flash FWD& Operating &  PWFA &  e-, 1.5 GeV, beam quality\\
 \hline\hline
\end{tabular}
\end{table}

Table~\ref{tab:facilities} highlights operating or upcoming facilities capable of delivering beams in the multi-GeV range or above, and which are essential tools to address crucial questions relevant to ALIC development in the next 5 to 10 years.
 During this period, the facilities already operating or in their building phase will be mainly dedicated to the development of a single stage of ALIC. %
However, \textbf{multi-stage challenges with high-energy beams }also need to be addressed. %
It is thus clear that \textbf{in the longer term, a facility to test staging with collider-like quality beams is necessary. }%

 The development of simulation tools  needed for the design of an advanced multi-TeV collider requires robust and sustained team efforts. Teams will consist of individuals with various skills and responsibilities: code developers, maintainers, and users. The sharing of modules/codes, interoperability via the development of libraries of algorithms and physics modules, as well as the definition of standards for simulation input/output and for data structures, should be encouraged.

\end{document}

%% file: shortphysicsforALEGRO.tex



\input mymacros.tex



While the Standard Model (SM) seems to describe all
observations from high-energy experiments, this
model is manifestly incomplete in many respects.   Experiments at the
LHC have emphasized this problem.   They have discovered the Higgs
boson, the last
particle predicted by the SM, but thus far, they have not given clues
to new physics beyond the SM. 

This situation calls for a number of measurements at e-/e+/gamma colliders that have the potential, first,
to break the impasse toward the existence of physics beyond the SM and, second, to learn the nature
of the new interactions. We list the most important of these below. The goals 1 to 6 have been described in studies for colliders in the
hundreds of GeV energy range---CEPC~\cite{CEPCpre}, FCC-ee~\cite{dEnterria:2016sca}, ILC~\cite{Baer:2013cma}, and CLIC~\cite{Linssen:2012hp}. However, it is also likely
that much important information about the nature of physics beyond the SM will lie beyond the reach
of these machines, at tens of TeV in the parton-parton center of mass frame. The goals 7 to 11 will require the ALIC collider 
that is the long-term goal of this initiative.


Here is a summary of the most important physics goals yet to be
realized at e$^-$/e$^+$ and $\gamma\gamma$ colliders.   A more detailed
discussion of each can be found in the Addendum~\cite{Addendum}. %
We emphasize that the last four goals can only be reached at high center of mass energies, i.e., with ANAs. %
\begin{enumerate}

\item  High-precision study of the $Z$ resonance and high-precision
  measurement of the $W$ mass, resolving current
  tensions among the precision electroweak measurements and testing
  the SM at the  $10^{-4}$ level.

\item Model-independent measurement of the Higgs boson
  couplings to 1\% precision.   This accesses deviations
  from SM model predictions at the level at which  effects of beyond-SM interactions
  would be visible. 

\item Search for invisible or exotic decays of the Higgs boson to the
  parts-per-mil level of branching fraction.

\item Measurement of the top quark electroweak form factors
  to parts per mil precision.   This accesses deviations
  from SM model predictions at the level at which  effects of beyond-SM interactions
  would be visible. 

\item Search for invisible particles pair-produced in e$^-$/e$^+$
  collisions.   An important objective is the pure Higgsino dark
  matter candidate, which would have a mass of 1~TeV.

\item Search for additional electroweak gauge bosons and signals of
  lepton and quark compositeness.  A 3~TeV e$^-$/e$^+$ collider would be
  sensitive to new bosons at 15~TeV and compositeness scales of
  60-80~TeV, far beyond the LHC capabilities.

\item  Search for pair-production of any new particles with multi-TeV
  masses that couple to
  the electroweak interactions.

\item Search for ``thermalization'' of Higgs boson production, the
  production of events with hundreds of $W$, $Z$, and Higgs bosons at
  center of mass energies above 10~TeV.

\item Exploration of the resonances of the new strong interactions
  associated with composite Higgs boson models.  These resonances are
  expected to appear above 10~TeV in the center of mass.

\item Determination of the geometry of extra space dimensions from the
  systematics of observed Kaluza-Klein resonances.   Given current
  constraints, e$^-$/e$^+$ or $\gamma\gamma$ 
 experiments above 20~TeV would be needed to draw firm conclusions.

\item Characterization of leptoquark bosons proposed to explain
  suggested anomalies in flavor physics, or other new particles that
  could be involved in explaining the systematics of flavor
  interactions.

\end{enumerate} 

For all of these goals, and more, we need precision experiments with
e$^-$/e$^+$ and $\gamma\gamma$ 
colliders and new experiments
that can access the TeV and 10~TeV energy scales with electroweak
probes.  It is thus important to have a long-term program to realize
these capabilities by inventing new technologies for electron and
positron acceleration. 


%% file: mymacros.tex


\def\beq{\begin{equation}}
\def\eeq#1{\label{#1}\end{equation}}
\def\eeqn{\end{equation}}


\newenvironment{Eqnarray}%
   {\arraycolsep 0.14em\begin{eqnarray}}{\end{eqnarray}}
\def\beqa{\begin{Eqnarray}}
\def\eeqa#1{\label{#1}\end{Eqnarray}}
\def\eeqan{\end{Eqnarray}}
\def\CR{\nonumber \\ }


\def\leqn#1{(\ref{#1})}






\def\st{\scriptstyle}
\def\sst{\scriptscriptstyle}

\def\overbar#1{\overline{#1}}
\let\littlebar=\bar
\let\bar=\overbar

\def\subscr#1{{\mbox{\scriptsize #1}}}



\def\etal{{\it et al.}}
\def\ie{{\it i.e.}}
\def\eg{{\it e.g.}}



\def\VEV#1{\left\langle{ #1} \right\rangle}
\def\bra#1{\left\langle{ #1} \right|}
\def\ket#1{\left| {#1} \right\rangle}
\def\vev#1{\langle #1 \rangle}



\def\lsim{\mathrel{\raise.3ex\hbox{$<$\kern-.75em\lower1ex\hbox{$\sim$}}}}
\def\gsim{\mathrel{\raise.3ex\hbox{$>$\kern-.75em\lower1ex\hbox{$\sim$}}}}

\def\Im{{\rm Im}}
\def\Re{{\rm Re}}


\def\D{{\cal D}}
\def\L{{\cal L}}
\def\M{{\cal M}}
\def\O{{\cal O}}
\def\W{{\cal W}}
\def\L{{\cal L}}



\def\One{{\bf 1}}
\def\hc{{\mbox{\rm h.c.}}}
\def\tr{{\mbox{\rm tr}}}
\def\half{\frac{1}{2}}
\def\thalf{\frac{3}{2}}
\def\third{\frac{1}{3}}
\def\tthird{\frac{2}{3}}

\def\del{\partial}
\def\Dslash{\not{\hbox{\kern-4pt $D$}}}
\def\dslash{\not{\hbox{\kern-2pt $\del$}}}

\def\Dlr{\mathrel{\raise1.5ex\hbox{$\leftrightarrow$\kern-1em\lower1.5ex\hbox{$D$}}}}



\def\Pl{{\mbox{\scriptsize Pl}}}
\def\eff{{\mbox{\scriptsize eff}}}
\def\CM{{\mbox{\scriptsize CM}}}
\def\GUT{{\mbox{\scriptsize GUT}}}
\def\BR{\mbox{\rm BR}}
\def\ee{e^+e^-}
\def\sstw{\sin^2\theta_w}
\def\cstw{\cos^2\theta_w}
\def\mz{m_Z}
\def\gz{\Gamma_Z}
\def\mw{m_W}
\def\mt{m_t}
\def\gt{\Gamma_t}
\def\mh{m_h}
\def\gmu{G_\mu}
\def\GF{G_F}
\def\alphas{\alpha_s}
\def\msb{{\bar{\scriptsize M \kern -1pt S}}}
\def\lmsb{\Lambda_{\msb}}
\def\drb{{\bar{\scriptsize D \kern -1pt R}}}
\def\ELER{e^-_Le^+_R}
\def\EREL{e^-_Re^+_L}
\def\ELEL{e^-_Le^+_L}
\def\ERER{e^-_Re^+_R}
\def\eps{\epsilon}



\def\spa#1#2{\langle #1 #2 \rangle}
\def\spb#1#2{[ #1 #2 ]}
\def\apb#1#2#3{\langle #1 #2 #3 ]}
\def\bpa#1#2#3{[ #1 #2 #3 \rangle}



\def\ch#1{\widetilde\chi^+_{#1}}
\def\chm#1{\widetilde\chi^-_{#1}}
\def\neu#1{\widetilde\chi^0_{#1}}
\def\s#1{\widetilde{#1}}


\makeatletter
\def\section{\@startsection{section}{0}{\z@}{5.5ex plus .5ex minus
 1.5ex}{2.3ex plus .2ex}{\large\bf}}
\def\subsection{\@startsection{subsection}{1}{\z@}{3.5ex plus .5ex minus
 1.5ex}{1.3ex plus .2ex}{\normalsize\bf}}
\def\subsubsection{\@startsection{subsubsection}{2}{\z@}{-3.5ex plus
-1ex minus  -.2ex}{2.3ex plus .2ex}{\normalsize\sl}}

\renewcommand{\@makecaption}[2]{%
   \vskip 10pt
   \setbox\@tempboxa\hbox{\small #1: #2}
   \ifdim \wd\@tempboxa >\hsize     
       \small #1: #2\par          
     \else                        
       \hbox to\hsize{\hfil\box\@tempboxa\hfil}
   \fi}

 \def\citenum#1{{\def\@cite##1##2{##1}\cite{#1}}}
 
\newcount\@tempcntc
\def\@citex[#1]#2{\if@filesw\immediate\write\@auxout{\string\citation{#2}}\fi
  \@tempcnta\z@\@tempcntb\m@ne\def\@citea{}\@cite{\@for\@citeb:=#2\do
    {\@ifundefined
       {b@\@citeb}{\@citeo\@tempcntb\m@ne\@citea\def\@citea{,}{\bf ?}\@warning
       {Citation `\@citeb' on page \thepage \space undefined}}%
    {\setbox\z@\hbox{\global\@tempcntc0\csname b@\@citeb\endcsname\relax}%
     \ifnum\@tempcntc=\z@ \@citeo\@tempcntb\m@ne
       \@citea\def\@citea{,}\hbox{\csname b@\@citeb\endcsname}%
     \else
      \advance\@tempcntb\@ne
      \ifnum\@tempcntb=\@tempcntc
      \else\advance\@tempcntb\m@ne\@citeo
      \@tempcnta\@tempcntc\@tempcntb\@tempcntc\fi\fi}}\@citeo}{#1}}
\def\@citeo{\ifnum\@tempcnta>\@tempcntb\else\@citea\def\@citea{,}%
  \ifnum\@tempcnta=\@tempcntb\the\@tempcnta\else
  {\advance\@tempcnta\@ne\ifnum\@tempcnta=\@tempcntb \else\def\@citea{--}\fi
    \advance\@tempcnta\m@ne\the\@tempcnta\@citea\the\@tempcntb}\fi\fi}
\makeatother